%% file: main.tex
\documentclass[10pt,twocolumn]{IEEEtran} 
\usepackage[noadjust]{cite}
\usepackage{pgfplots}
\usepackage{multicol}
\usepackage{multirow}
\pgfplotsset{width=4.7cm,compat=1.3}
\definecolor{mygr}{HTML}{A8A8A8}
\definecolor{mygrn}{HTML}{D3D3D3}
\usepackage{mwe}
\usepackage{scalerel}
\usepackage{balance}
\usepackage{amsmath,amssymb,amsfonts}
\usepackage{graphicx}
\usepackage{tikz}
\usepackage{tikz-qtree}

\def\BibTeX{{\rm B\kern-.05em{\sc i\kern-.025em b}\kern-.08em
    T\kern-.1667em\lower.7ex\hbox{E}\kern-.125emX}}

\begin{document}

\title{{\bf{\emph{CARE}}}: Lightweight Attack Resilient Secure Boot Architecture \\ with Onboard Recovery for RISC-V based SOC} 

\author{\large Avani Dave, Nilanjan Banerjee, Chintan Patel \\ 
 Computer Science \& Electrical  Engineering Department \\
 University of Maryland Baltimore County\\
E-mail: daveavani@umbc.edu}


\maketitle
\thispagestyle{empty}\pagestyle{empty}

\begin{abstract}
Recent technological advancements have proliferated the use of small embedded devices for collecting, processing, and transferring the security-critical information. The Internet of Things (IoT) has enabled remote access and control of these network-connected devices. Consequently, an attacker can exploit security vulnerabilities and compromise these devices. In this context, the secure boot becomes a useful security mechanism to verify the integrity
and authenticity of the software state of the devices. However, the current secure boot schemes focus on detecting the presence of potential malware on the device but not on disinfecting and restoring the software to a benign state. This manuscript presents \emph{CARE} - the first secure boot framework that provides detection, resilience, and onboard recovery mechanism for the compromised devices. The framework uses a prototype hybrid {\bf{\emph{CARE}: Code Authentication and Resilience Engine}} to verify the software state and restore it to a benign state. It uses Physical Memory Protection (PMP) and other security enchaining techniques of \mbox{RISC-V} processor to provide resilience from modern attacks. The state-of-the-art comparison and performance analysis results indicate that the proposed secure boot framework provides promising resilience and recovery mechanism with very little (8\%) performance and resource overhead. 
\end{abstract}

\input{Introduction}
\input{background}

\input{care}
\input{evaluation}
\input{discussion}

\input{conclusion}



\bibliographystyle{IEEEtran}
\bibliography{references}


%

%

\end{document}

%% file: Introduction.tex
\section{\bf{Introduction}}
\label{Intro}
\par The recent technological advancement has catastrophically increased the utilization of small embedded and IoT devices in applications ranging from industrial control systems, distributed sensing and actuation, vehicular and home automation systems. This increased utilization and inter-connectivity for collecting, transferring, and processing the security-critical information has made the small embedded and IoT devices, attractive targets for attacks. Prominent examples are the Rootkit \cite{Rootkit:2019}, the bios and secure boot attacks \cite{furtak:2014}, the Stuxnet \cite{Stuxnet}, and the Jeep hack \cite{jeep}. By-enlarge such attacks modify the targeted device's software state to leak, steal, tamper, or misuse the security-critical information for malicious activities. Such attacks can render the device into an unusable state. These types of attacks are commonly referred to as malicious code modification attacks or malware infestation. \par Secure boot process verifies the integrity and authenticity of devices' software state during boot time and ensures that the device boots-up with a known good code. Several secure boot methods have been proposed based on hardware \cite{Haj:2019, NSA}, software \cite{Sanctum:2018,Wong:2018}, and hardware/software co-design \cite{Tpm:2010, mcs:2015, Ope:2019, keystone}. While prior secure boot techniques focus on the detection of malicious code modification attacks, the problem of disinfecting the affected devices has been totally overlooked. The conventional device needs over-the-air or manual code re-flash to restore its normal operational state. A smart attacker can fail over-the-air code re-flash by corrupting the networking stack. This necessitates manual intervention. Sometimes the manual re-flash becomes relatively difficult due to the placement (in home security sensors and cameras, industrial and automotive control systems, ships) of the devices. \par To bridge this gap, the proposed work presents \emph{CARE} - first lightweight secure boot framework that provides detection, resilience, and onboard recovery mechanism for small embedded and IoT devices. 
\begin{figure}[h]
	\centering
	\includegraphics[width=\linewidth]{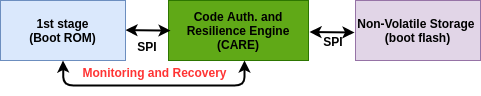}
	\caption{depicts the placement of the custom \emph{CARE} module (placed in between the first stage boot code (ROM) and second stage boot flash) that monitors the code integrity and authenticity during secure boot, and restores the corrupted flash memory region with the golden recovery data from secure ROM.}
	\label{fig: CARE}
	\vspace{-1em}
\end{figure}
\par Fig~\ref{fig: CARE} provides a high-level design overview of the proposed secure boot system. The hybrid \emph{CARE} module has two key components, namely, Code Authentication (CA) unit and Resilience Engine (RE). The framework measures the integrity and authenticity of the flash image using the CA sub-module. It triggers the RE sub-module upon detecting malicious code presence, else it continues the subsequent boot process. RE identifies and exclusively re-flashes only the corrupted flash memory regions with known good code from secure (backup) ROM. The framework disables unauthorized write and code execution from RAM by applying access control policies using the Physical Memory Protection (PMP) mechanism of the \mbox{RISC-V} processor. The framework performs integrity and authenticity check again and continues the secure boot process. This method ensures that irrespective of any malicious code modification attacks, the device will self recover and always boot up with a known good code. \\ The design, implementation, and evaluation of the proposed framework provides the following research contributions: 
\begin{itemize}
	\item{\bf{Code Integrity and Authenticity Measurement (CA) Tool}:} It demonstrates the lightweight implementation of integrity and authenticity measurement tools by reusing the same underlying hardware cryptographic core.
	\item{\bf{Resilience Engine (RE)}:} It demonstrates the first implementation of onboard resilience and recovery engine for small embedded and IoT devices. 
	\item{\bf{Lightweight Secure Boot Architecture}:} It provides FPGA prototype implementation of a lightweight, secure boot framework \emph{CARE} for small embedded and IoT devices. It enhances the attack resilience and security of the system by leveraging Side-Channel Analysis (SCA) and fault injection attack protection features of the \mbox{RISC-V} processor.  
\end{itemize}

%% file: background.tex
\section{\bf{Background and Related Work}}
Arbaugh et al. has proposed the first secure boot mechanism {\cite{Arbaugh:1997}, which measures the integrity of the system by verifying the integrity of the boot software code(stages). It performs a measured boot in which every stage verifies the subsequent stage's integrity before it gets executed. Authenticated boot verifies that software running on the system is coming from an authorized vendor. The Unified Extensible Firmware Interface (UEFI) specification since version 2.2 \cite{NSA} defines secure boot as a process to verify the integrity of each stage of the boot process by digest computation and comparing the result with a cryptographic signature. It requires access to a trustworthy public key database to verify the signature. The majority of the previous implementation of the secure boot systems performs either measured or authenticated boot, and very few perform both.\\
	\label{relcons1} 
	One of the popular methods for the secure boot is to use a discrete co-processor called the Trusted Platform Module (TPM) \cite{Tpm:2010}. TPM has a special purpose registers called Platform Configuration Registers (PCRs), which cannot be overwritten. PCR's can only be extended by hashing the software measurements together with the previous values of PCR. TPM can sign the PCRs with a private attestation key to generate a piece of attestation evidence. However, TPM is not suitable for small embedded or IoT devices due to space, size, and cost constraints. Intel's processor supports two modes of the secure boot - measured and verified and uses microcode as root-of-trust \cite{BootInt}. For measured boot, it uses TPM, and for verified boot, each component is signed by the manufacturer's key, and signatures are verified before loading the component. Microsoft's fTPM \cite{mcs:2015} provides a use-case of Arm TrustZone based secure boot and attestation. \mbox{RISC-V} based Sanctum \cite{Sanctum:2018} uses software-based secure boot and remote attestation. SMART \cite{Wong:2018} provides dynamic root-of-trust architecture for low-end devices. Keystone \cite{keystone} showcases a use case of Trusted Execution Environment (TEE) with enclaves. Haj et al. \cite{Haj:2019} presents lightweight hardware-based secure boot architecture for \mbox{RISC-V} based SoC. Google's recent open-source root-of-trust project Opentitan \cite{Ope:2019}, provides a sample implementation of secure root-of-trust. \par However, none of the available solutions have a secure recovery mechanism. Recently implementation Healed \cite{Healed:2019} and \cite{Secerase:2010} demonstrates recovery mechanisms. however, they both lack in proper secure boot implementation. To the best of our knowledge, the proposed work is the first implementation of a \emph{lightweight secure boot architecture with onboard resilience and recovery engine} for small embedded and IoT devices. 
	\vspace{-1em}
	\section{\bf{Adversarial Model \& Protection Axioms}}\label{themo}
	\par The proposed system assumes that the adversary can control the entire software code and data. The adversary can modify any writable memory and read memory region that is not protected by access control policies (using PMP) and secureIbex hardware features. The adversary can re-locate malware from one memory segment to another for hiding it from being detected. It also has full control over all Direct Memory Access (DMA) to access the main memory directly (e.g., RAM, flash or ROM) without going through the processor core. \par The proposed solution blocks un-authorized read, write, and code execution triggered from non-secure flash memory (to ROM) by applying PMP access control rules. It also leverages the special security feature (secureIbex) of the Ibex processor to protect the device from data independent timing - side-channel attack and fault injection attacks.

%% file: care.tex
\section{\bf{SYSTEM OVERVIEW}}
\subsection{\bf{Architecture and Design Choices}}
\begin{figure}[h]
	\begin{center}
		\includegraphics[width=3.3in]{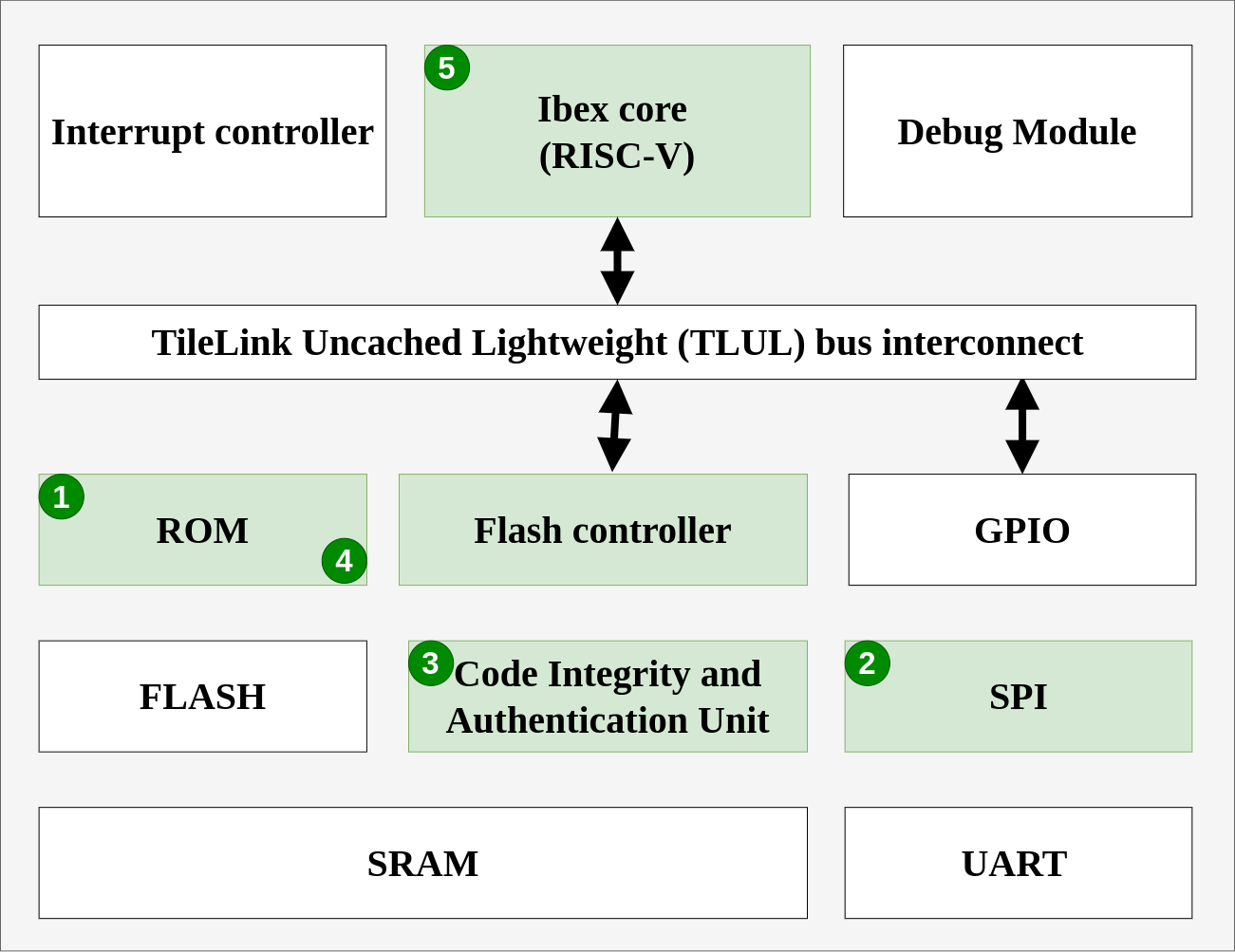}
	\end{center}
	\caption{Top-level design of \emph{CARE} - secure boot SoC On FPGA. Highlighted (light-green) boxes represents the Trusted Computing Base (TCB) components for the proposed framework.  
		\label{fig:soc}}
\end{figure}

The proposed secure boot system is built upon the Ibex \cite{Ibex:2018} \mbox{RISC-V} processor. The system is equipped with hardware-accelerated code integrity and authentication measurement (CA) unit, recovery engine (RE), secure boot, secure memory (ROM), and dedicated SPI bus as shown in Fig~\ref{fig:soc} ({\includegraphics[width=.3cm,height=.3cm]{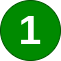}} to {\includegraphics[width=.3cm,height=.3cm]{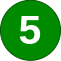}} highlights the key components).
 Notice that secure storage ROM has numbers {\includegraphics[width=.3cm,height=.3cm]{figs/n1}} and {\includegraphics[width=.3cm,height=.3cm]{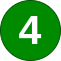}}, because it is used to store secure signing key, device information, and the recovery data. The flash controller module is used to translate read, erase, and program requests to low-level protocol signaling and timing. 
The proposed framework has incorporated the following security enhancing design features: 
\label{features}
{\includegraphics[width=.3cm,height=.3cm]{figs/n1}} \label{SS} {\bf{Secure Storage}} ROM is used for storing the device information such as vendor ID, Unique Device Identification (UUID), firmware revision, symmetric cryptographic shared key ($K$), and trusted recovery image.
{\includegraphics[width=.3cm,height=.3cm]{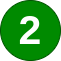}} {\bf{Secure SPI}} bus is used for communication between the ROM, flash, and \emph{CARE} module to protect the device from attacks launched using a shared internal bus \cite{Samebus:2017}. However, if the SPI tool's hardware or software gets corrupted, it can render and transfer incorrect or corrupted flash data. Therefore, the proposed design divides the flash image into 1~KB frame/chunks. Section~\S{\ref{frame}} covers the details of the frame data structure. Note that, this design choice is used for proof of concept implementation only and user can parameterize it to optimize the system performance.
{\includegraphics[width=.3cm,height=.3cm]{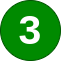}} \label{CAe}{\bf{Code Integrity \& Authentication (CA) Unit}} is implemented by reusing the same underlying hardware lightweight cryptographic-core (HMAC-SHA256), which performs both integrity and authenticity checks. This hardware reuse makes the proposed framework lightweight and resource-efficient, suitable for small embedded and IoT devices. {\includegraphics[width=.3cm,height=.3cm]{figs/n4}}\label{RES} {\bf{Resilience Engine (RE)}} is implemented in software and it re-flashes the affected flash memory region during the secure boot. It applies access control policies to protect the device from future attacks. {\includegraphics[width=.3cm,height=.3cm]{figs/n5}}\label{RES} {\bf{Ibex Core}} The Ibex core provides memory protection and access control using PMP. It also provides resilience from fault injection, data independent timing attacks by leveraging secureIbex feature. It inserts dummy instructions (such as NOP) at random interval to protect the system from side channel attacks. It performs ECC checking of flash blocks to protect the system from fault-injection attacks. 
\label{operation}
\subsection{\bf{System Operation}}
\par
\begin{figure}[h]
	\begin{center}
		\vspace{-1em}
		\includegraphics[width=3.3in]{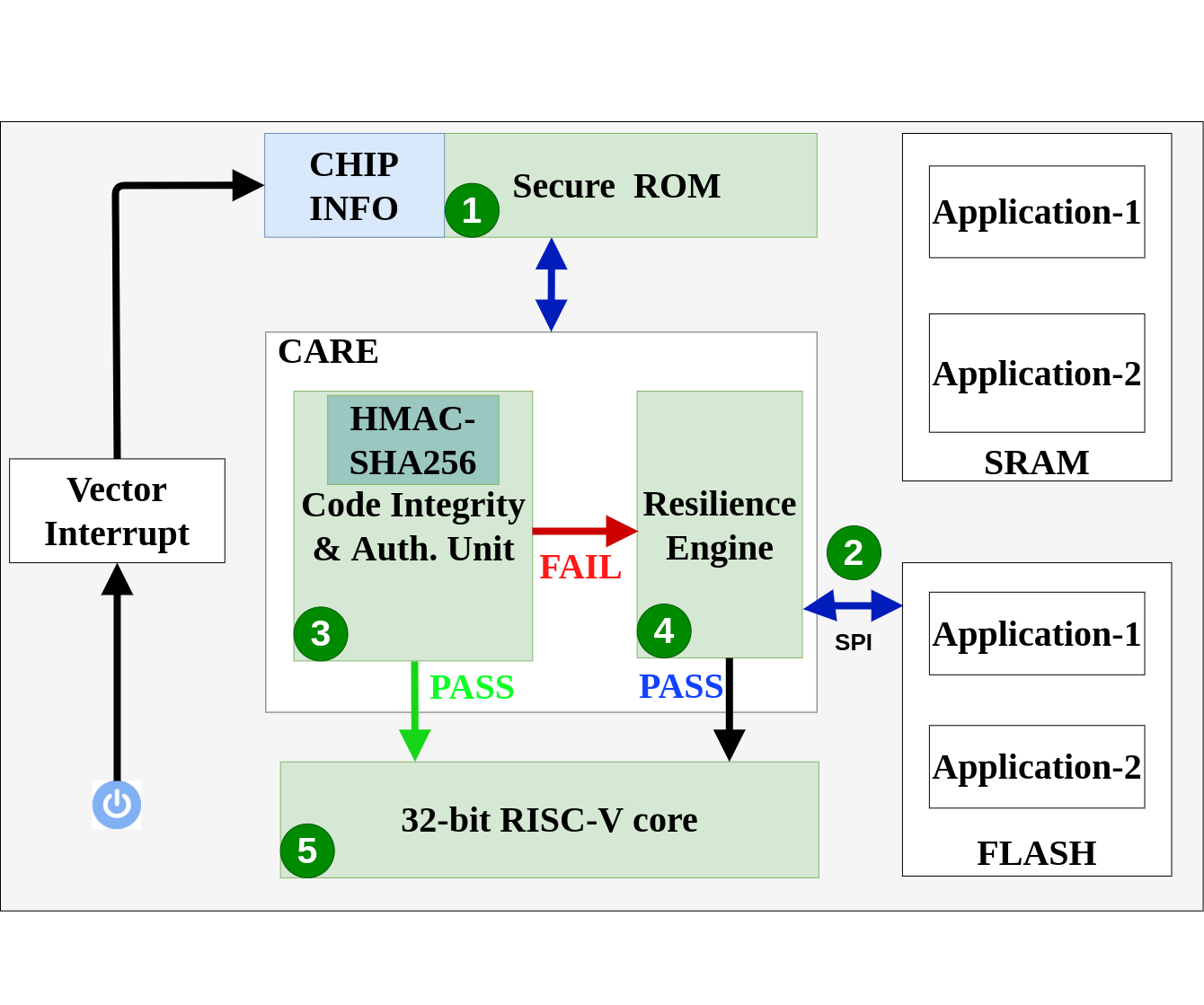}
	\end{center}
	\vspace{-2em}
	\caption{Shows the architecture design of the proposed framework. The pass arrows indicate that only the known good code will be passed to the \mbox{RISC-V} processor core for execution in any given case.
		\label{fig:care}}	\vspace{-1em}
\end{figure}

\begin{figure}[h]
	\begin{center}
		\includegraphics[width=3.3in]{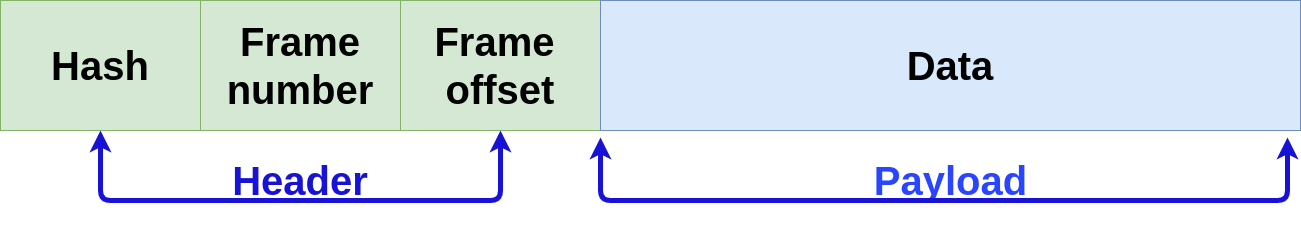}
	\end{center}
	\vspace{-1em}
	\caption{Frame data structure.    
		\label{fig:frame}}
\end{figure}

 The architecture design of the proposed framework is shown in Fig~\ref{fig:care} and the system operation is divided into three main steps: (1) System Initialization; (2) Code Integrity and Authenticity Check (Bootstrapping); and (3) Resilience Engine (RE). 
\label{design}
 \\
{{\bf1) System Initialization:}} Upon power-on, the system locates and executes the First Stage Boot Loader (FSBL) code from secure ROM to initialize the SPI and flash controllers. It then applies memory protection (using PMP) rules, and blocks un-authorized code read, write, and execution initiated from unprotected flash memory. It reads the chip information such as - device UUID, board version, and symmetric share key, generates derived keys, and hands off the control to the second stage boot code called the bootstrap.\\
{\bf2) {Bootstrapping System:}} \label{frame}
The secure boot starts with the bootstrapping process. It can be triggered by the hardware reset, power-on, or triggering from the external host via the General Purpose Input Output (GPIO) pin seven in the proposed framework. When the bootstrap is activated, the executable flash image is broken down into 1~KB frame chunks and sent sequentially to the host over the SPI bus. Each frame consists of a header and associated payload, as illustrated in Fig~\ref{fig:frame}. The header contains the signed digest of the frame data. The offset location field indicates the flash memory location, which will be used for code re-flash. The payload contains 968 bytes of the data for each frame.
\par The proposed framework has leveraged Hashed based Message Authentication Code (HMAC)'s  feature HMAC-SHA256 for signing the data and perform signature verification and SHA256 for digest computation. Few APIs were developed to reuse the same underlying cryptographic hardware HMAC-SHA256. The HW-SHA256 module computes the digest of each 1KB frame and compares it with the pre-computed hash for an integrity check. The HMAC-SHA256 uses a derived key to sign the computed digest, matches it with the "Hash" field in the frame header for authenticity check. The framework follows the same bootstrapping process for each subsequent frame. If everything passes, the device boots up with verified code, else it triggers the resilience engine.\\
{\bf3) {Resilience Engine (RE):}}
The resilience engine acts as follows: (1) It identifies the frame number and offset location of the corrupted frame, and locates the corresponding golden frame data from the secure EEPROM. (2) It exclusively re-flashes the corrupted flash memory region with the known good code. (3) It locks un-authorized read-write access to the memory using the PMP mechanism. These steps ensure that the proposed device will always boot up with a known good code irrespective of any memory modification attacks.  
\vspace{-1em}

%% file: evaluation.tex
\section{\bf{Evaluation}}
This section describes the chain-of-trust theory, resource utilization, and performance analysis for each sub-module and overall framework in the proposed framework, and presents state-of-the-art comparison results.  
\subsection{\bf{Chain-of-Trust}}
The proposed framework breaks down the entire flash image into 1~KB chunks/frames and measures the code's integrity and authenticity at the frame level. The following equation represents the chain-of-trust: 
\[ 
V_0{} = True
\]
\vspace{-1.5em}
\begin{equation}
V\textsubscript{\emph{i}+1} = V_\emph{i}{}\;\;\;\; \& \;\;\;\;                                               S_\emph{fi}{}(f\textsubscript{\emph{i}+1} )\;\;\;\; \& \;\;\;\;                                               I_\emph{fi}{}(f\textsubscript{\emph{i}+1} )
\end{equation}
$V_\emph{i}{}$ is the boolean value representing the software state verification (both integrity and authenticity) of the $i$th frame, and \& is the boolean AND operation. $I_\emph{fi}{}$ and $S_\emph{fi}{}$ represents the integrity and signature (authenticity) verification functions respectively. $I_\emph{fi}{}$ takes frame data (f\textsubscript{\emph{i}+1}) as input argument, performs a cryptographic hash, compares the result with golden digest value, and returns a boolean result. The $S_\emph{fi}{}$ signs the digest, matches with frame header value, and gives a boolean result. The following equation~\ref{eq2} calculates the estimated increase in boot time (T\textsubscript{$\Delta$}) for proposed secure boot with \emph{CARE}. 
\vspace{-1em}
\begin{equation}\label{eq2}
T\textsubscript{$\Delta$} = t\textsubscript{fm\_0}(I_0{}(f_1{}) + S_0{}(f_1{}))\; +\; t\textsubscript{fm}\sum_{\emph{i}=1}^{n}[I_\emph{i}{}(f_{\emph{i}+1}{}) + S_\emph{i}{}(f_{\emph{i}+1}){}]
\end{equation}
Where t\textsubscript{fm\_0} and t\textsubscript{fm} are the execution time for the first and all other frames, respectively. By design, the framework, first matches the frame number of the received frame and clears the flash region to re-flash it with a trusted code. Therefore, the first frame processing requires more time than the remaining frames. In case of verification failure of any frame, the framework triggers RE sub-module, which intern re-flashes the corrupted frame with know good recovery data. This process increases the boot time by a fraction, as discussed in subsection~{\S\ref{PA2}}. The system analysis is carried out by evaluating performance, hardware-software resource utilization, and energy consumption of the CA unit, RE, and overall system.    
\subsection{\bf{Code Integrity and Authentication (CA) Unit}}\label{PA}
The cryptographic-core (HMAC-SHA256) is the key component of the Code Integrity and Authentication (CA) unit. The test setup first uses both hardware and software \cite{swcore:2019} implementation of cryptographic-core running on FPGA for performance evaluation, as shown in Table~1. The system computes the digest of 256 Bytes of data for performance and energy efficiency evaluation. 
\begin{table}[h]\label{tab:hwswperf}
	\caption{Performance Analysis of Crypto-core on FPGA.}
	\vspace{-1em}
	\scalebox{1.08}[1.0]{\tt
		\begin{tabular}{@{}lcc@{}}
			\hline
			{Parameters} & {Software} & Hardware \\ \hline
			\multicolumn{1}{l|}{Cycles~(c)} & \multicolumn{1}{c|}{\bf{47033}}& {\bf{2926}} \\
			\multicolumn{1}{l|}{Frequency~(F)~(MHz)} & \multicolumn{1}{c|}{100} & 100    \\
			\multicolumn{1}{l|}{Block~(b)} & \multicolumn{1}{c|}{256}& 256    \\
			\multicolumn{1}{l|}{Throughput~(T)~(Mbps)}& \multicolumn{1}{c|}{.54} & 8.74   \\
			\multicolumn{1}{l|}{Time~($\mu$sec)} & \multicolumn{1}{c|}{470.33} & 29.26    \\
			\multicolumn{1}{l|}{Energy Consumption~(E)} & \multicolumn{1}{c|}{197.06} & 12.25 \\
			\multicolumn{1}{l|}{Energy Efficiency} & \multicolumn{1}{c|}{\bf{92.68}} & {\bf{0.358}} \\ \hline
	\end{tabular} }
\end{table}  
Table~1 shows the performance increase of 16x with 92${\%}$ less power utilization using hardware-based cryptographic-core. Furthermore, the proposed cryptographic-core is lightweight and consumes less energy than the recent state-of-the-art HMAC-SHA256 implementations, as depicted in Table~2. 
\begin{table}[h]\label{tab:quanthw} 
	\caption{Cross Comparison of Crypto-core on FPGA.}
	\vspace{-1em}
	\scalebox{.75}[1.0]{\tt
		\begin{tabular}{@{}lcccc@{}}
			\hline
			\multicolumn{1}{l|}{Work}              & \multicolumn{1}{c|}{\emph{Area}} & \multicolumn{1}{c|}{Freq.} & \multicolumn{1}{c|}{Energy Con.} & \multicolumn{1}{c}{Device} \\ \hline
			\multicolumn{1}{l|}{This work}         & \multicolumn{1}{c|}{2591}  & \multicolumn{1}{c|}{100} &  \multicolumn{1}{c|}{.012} & \multicolumn{1}{c}{Artix-7}  \\
			\multicolumn{1}{l|}{Opentitan \cite{Ope:2019}}         & \multicolumn{1}{c|}{2693}  & \multicolumn{1}{c|}{100} & \multicolumn{1}{c|}{.022} &  \multicolumn{1}{c}{Artix-7}  \\
			\multicolumn{1}{l|}{He et al.\cite{hmacsha256:2019}}         & \multicolumn{1}{c|}{7219}  & \multicolumn{1}{c|}{116.24} &  \multicolumn{1}{c|}{1.20} & \multicolumn{1}{c}{Arria II GX}   \\
			\multicolumn{1}{l|}{He et al.\cite{hmacsha256:2019}}         & \multicolumn{1}{c|}{10918}  & \multicolumn{1}{c|}{87} &  \multicolumn{1}{c|}{1.80} & \multicolumn{1}{c}{Arria II GX}   \\
			\multicolumn{1}{l|}{Juliato et al. \cite{Juliato2011FPGAIO}}      & \multicolumn{1}{c|}{2347}  & \multicolumn{1}{c|}{138.10} &   \multicolumn{1}{c|}{.48} & \multicolumn{1}{c}{Stratix III}    \\
			\multicolumn{1}{l|}{Juliato et al. \cite{Juliato:2010}}      & \multicolumn{1}{c|}{4281}  & \multicolumn{1}{c|}{67} &   \multicolumn{1}{c|}{.285} & \multicolumn{1}{c}{CycloneII}    \\
			\multicolumn{1}{l|}{Juliato et al. \cite{Juliato:2010}}      & \multicolumn{1}{c|}{6874}  & \multicolumn{1}{c|}{41.25} &   \multicolumn{1}{c|}{.431} & \multicolumn{1}{c}{CycloneII}    \\
			\hline
			\vspace{-1em}
		\end{tabular}
	}
\end{table}
As seen from Table~2, \cite {Juliato2011FPGAIO} requires a relatively low area but consumes high energy. \cite{Juliato:2010} presents base and DPA resilient cryptographic-core but uses more area and power. The work presented in \cite{hmacsha256:2019} optimizes the core for high throughput while compromising area and energy consumption, which makes both of them unsuitable for small embedded and IoT devices. The cryptographic-core used in the proposed work is an area and energy-optimized version of opentitan \cite{Ope:2019}. \\ 
\vspace{-1em}
\subsection{\bf{Resilience Engine (RE)}} 
\label{swres}
\par The Resilience Engine (RE) sub-module is implemented in software for the Proof-Of-Concept (POC) work. The test application of 5.6~KB is used for POC validation. RE requires {\bf{61}} additional lines of code (C language) for secure boot, and increases the secure ROM by 5~KB to store recovery data. The Resilience Engine (RE) requires 968~bytes of recovery data for every 1~KB of the flash image. To limit the size of the recovery data storage on the ROM, the system developer can select the necessary code modules for recovery to bring the system to a minimum working state. Although this feature is not implemented in the presented work due to small test applications. 

\subsection{\bf{System Performance}}\label{PA2}
\par The system divides a test application of 5.6~KB into six 1~KB frames and performs integrity and authenticity checks for the performance evaluation. The total boot-time and energy consumption with and without \emph{CARE} based secure boot is calculated for the test application running on FPGA. The timing analysis details are depicted in Table~\ref{tab:timeanalysis}. 
\begin{table}[h]
	\caption{Timing Analysis of Secure Boot on FPGA.}
	\vspace{-1em}
	\scalebox{.72}[1.0]{	\tt
		\begin{tabular}{@{}lcc@{}}
			\hline
			\multicolumn{1}{l|}{Parameters} & \multicolumn{1}{c|}{Without \emph{CARE}} & With \emph{CARE} \\ \hline
			\multicolumn{1}{l|}{Cycles req. for the first frame (c)} & \multicolumn{1}{c|}{553611} &  576083   \\
			\multicolumn{1}{l|}{Cycles~(rest of frames)~(c)} & \multicolumn{1}{c|}{103330}& 133790    \\
			\multicolumn{1}{l|}{Total Cycles~(C)}& 
			\multicolumn{1}{c|}{\bf{656941} } & \bf{709873}   \\
			\multicolumn{1}{l|}{Frequency~(F)~(MHz)}& \multicolumn{1}{c|}{100} & 100    \\
			\multicolumn{1}{l|}{Time~(t)~($\mu$sec)}& \multicolumn{1}{c|}{\bf{6569.41}}& \bf{7098.73} \\
			\multicolumn{1}{l|}{Energy Consumption~(E) }& \multicolumn{1}{c|}{\bf{2752.58}} & {\bf{2974.36}} \\ \hline
			Time difference D\textsubscript{$\Delta$} = {\bf{529.32}}~$\mu$sec \\ \hline
			\vspace{-2em}
		\end{tabular}
	}
	\label{tab:timeanalysis}
\end{table} 
The framework uses equation~\ref{eq2} to calculate the total execution time $T$. As explained earlier, the first frame requires more cycles and time. The rest of the frames consume an equal number of cycles. The secure boot with \emph{CARE} consumes 8$\%$ more energy and requires additional $D_{\Delta} = 529 \mu sec $ boot-time. The proposed RE sub-module requires an additional 334.475~$\mu$ sec to re-flash 968 bytes of data for each affected frame. This performance overhead (only 8$\%$ for the test application) is insignificant compared to the security and resilience, it provides.
\subsection{\bf{Comparison with the state-of-the-art solutions}}
\par The majority of the available secure boot implementations focus on detecting and preventing malicious code modification attacks. They generally stops the code execution or resets the system to protect it from attacks. These systems largely lack in providing recovery mechanism. Furthermore, our architecture was design using open-sourced \mbox{RISC-V} ISA, which is relatively new, and we did not found any secure boot implementation on \mbox{RISC-V} that provides the recovery mechanism. Therefore, we chose to compare the (quantitative) hardware footprint requirements of the recently proposed \mbox{RISC-V} based secure boot architectures with this work. The qualitative and quantitative comparison of the proposed secure boot framework with state-of-the-art solutions are presented in Table~\ref{qunt11} and Table~\ref{qant11}. 
{\bf{Qualitative Comparison}:} Table~\ref{qunt11} shows that \emph{CARE}, \cite{Ope:2019} and \cite{ASiddi} are hybrid secure boot systems. All three uses cryptographic-core SHA256 for integrity checking. \emph{CARE} and \cite{Ope:2019} uses HMAC-SHA256, and \cite{ASiddi} uses AES for authenticity check. \cite{ASiddi} uses a discrete TPM module connected to FPGA for secure boot.
\begin{table}[h]
	\caption{Qualitative Comparison}\label{qunt11}
	\vspace{-1em}
	\scalebox{.75}[1.0]{\tt
		\begin{tabular}{@{}lcccc@{}} 
			\hline
			\multicolumn{1}{l|}{Parameters}        & \multicolumn{1}{c|}{\emph{CARE}} &
			\multicolumn{1}{c|}{Optitan\cite{Ope:2019}} &  \multicolumn{1}{c|}{Haj\cite{Haj:2019}} & \multicolumn{1}{c}{Ref.\cite{ASiddi}} \\ \hline
			\multicolumn{1}{l|}{Design Type}         & \multicolumn{1}{c|}{Hybrid}  & \multicolumn{1}{c|}{Hybrid} & \multicolumn{1}{c|}{HW} & \multicolumn{1}{c}{Hybrid} \\
			\multicolumn{1}{l|}{Secure boot function}         & \multicolumn{1}{c|}{hmacSha2}  & \multicolumn{1}{c|}{hmacSha2} & \multicolumn{1}{c|}{Sha3} &  \multicolumn{1}{c}{Sha2}   \\ 
			\multicolumn{1}{l|}{Rom for Secure boot}         & \multicolumn{1}{c|}{yes}  & \multicolumn{1}{c|}{yes} & \multicolumn{1}{c|}{yes} &  \multicolumn{1}{c}{no/TPM}   \\
			\multicolumn{1}{l|}{Integrity Check}         & \multicolumn{1}{c|}{yes}  & \multicolumn{1}{c|}{yes} & \multicolumn{1}{c|}{yes} &  \multicolumn{1}{c}{yes}  \\
			\multicolumn{1}{l|}{Authenticity Check}         & \multicolumn{1}{c|}{yes}  & \multicolumn{1}{c|}{no} & \multicolumn{1}{c|}{yes} &  \multicolumn{1}{c}{yes}    \\
			\multicolumn{1}{l|}{Recovery}         & \multicolumn{1}{c|}{yes}  & \multicolumn{1}{c|}{no} & \multicolumn{1}{c|}{no} &  \multicolumn{1}{c}{no}  \\ 
			\multicolumn{1}{l|}{Lightweight}         & \multicolumn{1}{c|}{yes}  & \multicolumn{1}{c|}{yes} & \multicolumn{1}{c|}{no} &  \multicolumn{1}{c}{no}  \\    \hline
		\end{tabular}
	}
\end{table}
\par Haj et al.\cite{Haj:2019} is a pure hardware-based secure boot with TEE and resource-heavy cryptographic-cores (ECDSA Table~\ref{qant11} and sha3). Another implementation sanctum \cite{Sanctum:2018} provides a software-based secure boot by using secure enclaves. All three of them (\cite{ASiddi}, \cite{Haj:2019}, \cite{Sanctum:2018}) are resource heavy and not suitable for our targeted small embedded and IoT devices. Only \emph{CARE} and \cite{Ope:2019}
opentitan are lightweight solution. However,the baseline opentitan does not have support for cryptocore (HMAC-SHA256) for authenticity check, PMP, secureIbex register, and onboard recovery engine such as \emph{CARE}. In addition,  reuses the same hardware HMAC-SHA256 reuse for both integrity and authenticity check makes \emph{CARE} lightweight and suitable for our targeted devices.\\
{\bf{Quantitative Comparison}:}\label{quantify}
Table~\ref{qant11} enumerates the quantitative comparison of \emph{CARE}, \cite{Ope:2019} opentitan and Haj et al.\cite{Haj:2019} systems. Since \cite{ASiddi} uses discrete TPM module attached to FPGA for secure boot, the architecture design becomes different and heavy. Therefore, it is not suitable for quantitative analysis. 
\vspace{-1em}
\begin{table}[h]
	\caption{Qualitative comparison}\label{qant11}
	\vspace{-1em}
	\scalebox{.90}[1.1]{\tt
		\begin{tabular}{@{}lccccc@{}}\hline
			\multicolumn{2}{l|}{Parameters} & \multicolumn{3}{c}{FPGA Hardware}  \\ \hline
			\multicolumn{1}{l|}{Component}& \multicolumn{1}{c|}{Work} & \multicolumn{1}{c|}{LUTs} &
			\multicolumn{1}{c|}{Regs} &  \multicolumn{1}{c}{Cells} \\ \hline
			\multicolumn{1}{l|}{Complete SoC} &
			\multicolumn{1}{c|}{\emph{CARE}} & \multicolumn{1}{c|}{18620}  & \multicolumn{1}{c|}{8723} & \multicolumn{1}{c}{29792}    \\
			\multicolumn{1}{l|}{}   & 
			\multicolumn{1}{c|}{Opentitan\cite{Ope:2019}} &
			\multicolumn{1}{c|}{26468}  & \multicolumn{1}{c|}{1125} &  \multicolumn{1}{c}{42348.8}  \\
			\multicolumn{1}{l|}{}   & 
			\multicolumn{1}{c|}{Haj\cite{Haj:2019}} &
			\multicolumn{1}{c|}{N/A}  & \multicolumn{1}{c|}{N/A} &  \multicolumn{1}{c}{N/A} \\ \hline
			\multicolumn{1}{l|}{Crypto-Engine} &
			\multicolumn{1}{c|}{\emph{CARE}} & \multicolumn{1}{c|}{2591}  & \multicolumn{1}{c|}{1715} & \multicolumn{1}{c}{4145.6}   \\
			\multicolumn{1}{l|}{}   & 
			\multicolumn{1}{c|}{Opentitan\cite{Ope:2019}} &
			\multicolumn{1}{c|}{2693}  & \multicolumn{1}{c|}{1739} &  \multicolumn{1}{c}{4308.8} \\
			\multicolumn{1}{l|}{}   & 
			\multicolumn{1}{c|}{Haj\cite{Haj:2019}} &
			\multicolumn{1}{c|}{27170}  & \multicolumn{1}{c|}{6722} &  \multicolumn{1}{c}{43472} \\  \hline
	\end{tabular}}
\end{table}
Note that Haj et al.\cite{Haj:2019} does not provide complete secure boot SoC hardware foot-print. Therefore, that row has N/A - Not Available for LUTs, Regs, and Cells fields. 
\begin{figure}[!ht]
	\centering
	\begin{minipage}{.45\linewidth}
		\begin{tikzpicture}
		\begin{axis}[
		ybar stacked,
		bar width=25pt,
		nodes near coords,
		enlarge x limits=0.15,
		ymin=0,
		legend style={at={(0.5,-0.20)},
			anchor=north,legend columns=-1},
		symbolic x coords={CARE, Optitan, Haj},
		xtick=data,
		x tick label style={rotate=0,anchor=east, yshift= -7, xshift = 13 },
		axis lines*=left,
		] 
		\addplot+[draw=none, fill=mygr, ybar] plot coordinates {(CARE,9.53) (Optitan,9.91)  (Haj,90.47)};
		\end{axis}
		\end{tikzpicture}
		\caption{table}{($\%$) Hardware Overhead of Crypto-Engine }
		\label{img1}
	\end{minipage}
	\hspace{.05\linewidth}
	\begin{minipage}{.45\linewidth}
		\begin{tikzpicture}
		\begin{axis}[
		ybar stacked,
		bar width=35pt,
		nodes near coords,
		enlarge x limits=0.15,
		ymin=0,
		legend style={at={(0.5,-0.20)},
			anchor=north,legend columns=-1, yshift= -24},
		symbolic x coords={CARE, Opentitan},
		xtick=data,
		x tick label style={rotate=0,anchor=east, yshift= -7, xshift = 13 },
		axis lines*=left,
		] 
		\addplot+[draw=none, fill=mygr, ybar] plot coordinates {(CARE,70.34) (Opentitan,100)};
		\end{axis}
		\end{tikzpicture}
		\caption{table}{($\%$) Hardware Overhead for Complete SoC}
		\label{img2}
	\end{minipage}
\end{figure}
\begin{figure}[!ht]
	\centering
	\begin{minipage}{.45\linewidth}
		\begin{tikzpicture}
		\begin{axis}[
		ybar stacked,
		bar width=35pt,
		nodes near coords,
		enlarge x limits=0.15,
		ymin=0,
		legend style={at={(0.5,-0.20)},
			anchor=north,legend columns=-1},
		symbolic x coords={CARE, Opentitan},
		xtick=data,
		x tick label style={rotate=0,anchor=east, yshift= -7, xshift = 17},
		axis lines*=left,
		] 
		\addplot+[draw=none, fill=mygrn, ybar] plot coordinates {(CARE,12)  
			(Opentitan,12) };
		\addplot+[draw=none, fill=mygr,ybar] plot coordinates {(CARE,6) (Opentitan,)};
		\legend{\strut Opentitan, \strut CARE}
		\end{axis}
		\end{tikzpicture}
		\caption{table}{Memory Overhead in KB}
		\label{img3}
	\end{minipage}
	\hspace{.05\linewidth}
	\begin{minipage}{.45\linewidth}
		\begin{tikzpicture}
		\begin{axis}[
		ybar stacked,
		bar width=35pt,
		nodes near coords,
		enlarge x limits=0.15,
		ymin=0,
		legend style={at={(0.5,-0.20)},
			anchor=north,legend columns=-1, yshift= 0},
		symbolic x coords={Perf., Energy},
		xtick=data,
		x tick label style={rotate=0,anchor=east, yshift= -7, xshift = 17 },
		axis lines*=left,
		] 
		\addplot+[draw=none, fill=mygrn, ybar] plot coordinates {(Perf.,92) (Energy,92)};
		\addplot+[draw=none, fill=mygr, ybar] plot coordinates {(Perf.,8) (Energy,8)};
		\legend{\strut Opentitan, \strut CARE}
		\end{axis}
		\end{tikzpicture}
		\caption{table}{($\%$) Performance and Energy Consumption Overhead}
		\label{img4}
	\end{minipage}
\end{figure}
\par Table~\ref{img1} shows percentage hardware overhead of the cryptographic-core for all three solutions. (Note: "Optitan" name is used for Opentitan representation only). The ECDSA core from Haj et al. requires 90$\%$ more hardware resources than HMAC from \emph{CARE}. In-fact, the area required by Haj et al. based cryptographic-core is {\bf{14x}} larger than that of \emph{CARE}'s cryptographic-core. The comparison of asymmetric and symmetric cryptographic hardware requirements provide an initial estimation of overall hardware overhead requirements. Additionally, Haj et al. \cite{Haj:2019} requires two 64 bit RISC-V cores for Trusted Execution Environment (TEE) implementation, hardware SHA3 for hashing, and configurable LFSR-based Physical Unclonable Function (CoLPUF) for key generation, boot sequencer, and key management unit. These makes it a resource-heavy solution and unsuitable for small embedded and IoT devices. The percentage ($\%$) overhead of complete SOC is calculated between \emph{CARE} and opentitan \cite{Ope:2019} only, as Haj et al.\cite{Haj:2019} does not provide details of SoC hardware foot-print. Table~\ref{img2} depicts that the complete SoC with \emph{CARE} utilizes 29.66$\%$ less area. Table~\ref{img3} shows extended ROM region by 5~KB, and Table~\ref{img4} depicts performance and energy consumption overhead of 8$\%$ each compared with opentitan secure boot SoC. 

%% file: discussion.tex
\section{Discussion} 
\par A possible alternate of this work is a method that allows the device to boot from an SD card or a USB flash drive. However, providing physical security to an external card is extremely difficult. It can result in malicious actors replacing any hardware or software module in the system without security controls if the external card is lost or stolen. Second, the practicality of storing the golden image on EEPROM and not on the flash, as ROM's cost decreases ($< .50$ cents for 32~KB, when bought in bulk), making it an affordable alternative with more security. The \emph{CARE} based SoC design is limited to support for small embedded and IoT devices, which does not require frequent application code update. However, the design can use EEPROM to update the recovery image if the system needs it.

%% file: conclusion.tex
\section{Conclusion}
\par  This paper has presented a lightweight, secure boot framework with an onboard recovery and protection mechanism for small embedded and IoT devices, to protect it from malicious code modification attacks. It provides code modification attack detection, recovery, and prevention tools that assure the user that the device will always boot with a known good code. The framework achieves these by using a prototype \emph{CARE} module. It reuses the same cryptographic-core for authenticity and integrity check. The comparison of the proposed solution with the state-of-the-art secure boot implementations demonstrate that the proposed framework shows promising resilience and recovery methods with only 8$\%$ performance and energy consumption overhead and a minimal increase in hardware-software resource utilization.  